\documentclass[apj]{emulateapj}

\usepackage{apjfonts}
\usepackage{amssymb, amsmath}
\usepackage{natbib}
\usepackage{appendix}
\newcommand\degree{\degr}
\newcommand\degrees\degree

\DeclareSymbolFont{UPM}{U}{eur}{m}{n}
\DeclareMathSymbol{\umu}{0}{UPM}{"16}
\let\oldumu=\umu
\renewcommand\umu{\ifmmode\oldumu\else\math{\oldumu}\fi}
\newcommand\micro{\umu}
\renewcommand\micron{\micro m}
\newcommand\microns \micron

\let\oldsim=\sim
\renewcommand\sim{\ifmmode\oldsim\else\math{\oldsim}\fi}
\let\oldpm=\pm
\renewcommand\pm{\ifmmode\oldpm\else\math{\oldpm}\fi}
\newcommand\by{\ifmmode\times\else\math{\times}\fi}

\newbox{\wdbox}
\renewcommand\c{\setbox\wdbox=\hbox{,}\hspace{\wd\wdbox}}
\renewcommand\i{\setbox\wdbox=\hbox{i}\hspace{\wd\wdbox}}

\newcount\timect
\newcount\hourct
\newcount\minct
\newcommand\now{\timect=\time \divide\timect by 60
         \hourct=\timect \multiply\hourct by 60
         \minct=\time \advance\minct by -\hourct
         \number\timect:\ifnum \minct < 10 0\fi\number\minct}

\catcode`@=11

\newcommand\comment[1]{}
\newcommand\commenton{\catcode`\%=14}
\newcommand\commentoff{\catcode`\%=12}

\renewcommand\math[1]{$#1$}
\newcommand\mathshifton{\catcode`\$=3}
\newcommand\mathshiftoff{\catcode`\$=12}

\comment{the backslash is necessary}

\comment{alignment tab}

\let\atab=&
\newcommand\atabon{\catcode`\&=4}
\newcommand\ataboff{\catcode`\&=12}

\let\oldmsp=\sp
\let\oldmsb=\sb
\def\sp#1{\ifmmode
           \oldmsp{#1}%
         \else\strut\raise.85ex\hbox{\scriptsize #1}\fi}
\def\sb#1{\ifmmode
           \oldmsb{#1}%
         \else\strut\raise-.54ex\hbox{\scriptsize #1}\fi}
\newbox\@sp
\newbox\@sb
\def\sbp#1#2{\ifmmode%
           \oldmsb{#1}\oldmsp{#2}%
         \else
           \setbox\@sb=\hbox{\sb{#1}}%
           \setbox\@sp=\hbox{\sp{#2}}%
           \rlap{\copy\@sb}\copy\@sp
           \ifdim \wd\@sb >\wd\@sp
             \hskip -\wd\@sp \hskip \wd\@sb
           \fi
        \fi}
\def\msp#1{\ifmmode
           \oldmsp{#1}
         \else \math{\oldmsp{#1}}\fi}
\def\msb#1{\ifmmode
           \oldmsb{#1}
         \else \math{\oldmsb{#1}}\fi}
\def\supon{\catcode`\^=7}
\def\supoff{\catcode`\^=12}
\def\subon{\catcode`\_=8}
\def\suboff{\catcode`\_=12}
\def\supsubon{\supon \subon}
\def\supsuboff{\supoff \suboff}

\newcommand\actcharon{\catcode`\~=13}
\newcommand\actcharoff{\catcode`\~=12}

\newcommand\paramon{\catcode`\#=6}
\newcommand\paramoff{\catcode`\#=12}

\comment{And now to turn us totally on and off...}

\newcommand\reservedcharson{\commenton \mathshifton \atabon \supsubon \actcharon
	\paramon}

\newcommand\reservedcharsoff{\commentoff \mathshiftoff \ataboff
	\supsuboff \actcharoff \paramoff}

\catcode`@=12
\reservedcharsoff

\reservedcharson

\comment{ Must have ONLY ONE of these... trust these macros, they work

}

\newcommand{\squishlist}{
 \begin{list}{$\bullet$}
  { \setlength{\itemsep}{1pt}
     \setlength{\parsep}{0pt}
     \setlength{\topsep}{3pt}
     \setlength{\partopsep}{0pt}
     \setlength{\leftmargin}{2.0em}
     \setlength{\labelwidth}{1.5em}
     \setlength{\labelsep}{0.5em} } }

\newcommand{\squishend}{
  \end{list}  }

\reservedcharson
\actcharon

\shorttitle{New Analysis Indicates No Thermal Inversion in the Atmosphere of HD~209458\lowercase{b}}
\shortauthors{Diamond-Lowe {\em et al.}}

\begin{document}

\title{New Analysis Indicates No Thermal Inversion in the Atmosphere of HD~209458\lowercase{b}}

\author{Hannah\ Diamond-Lowe\altaffilmark{1}}
\author{Kevin B.\ Stevenson\altaffilmark{1,3}}
\author{Jacob L.\ Bean\altaffilmark{1}}
\author{Michael R. Line\altaffilmark{2}}
\author{Jonathan J. Fortney\altaffilmark{2}}
\affil{\sp{1}Department of Astronomy and Astrophysics, University of Chicago, 5640 S Ellis Ave, Chicago, IL 60637, USA\\ \sp{2}Department of Astronomy and Astrophysics, University of California, Santa Cruz, 1156 High Street, Santa Cruz, CA 95064, USA\\ \sp{3} NASA Sagan Fellow}

\email{E-mail: hdiamondlowe@uchicago.edu}

\begin{abstract}

An important focus of exoplanet research is the determination of the atmospheric temperature structure of strongly irradiated gas giant planets, or hot Jupiters. HD~209458b is the prototypical exoplanet for atmospheric thermal inversions, but this assertion does not take into account recently obtained data or newer data reduction techniques. We re-examine this claim by investigating all publicly available \textit{Spitzer Space Telescope} secondary-eclipse photometric data of HD~209458b and performing a self-consistent analysis. We employ data reduction techniques that minimize stellar centroid variations, apply sophisticated models to known \textit{Spitzer} systematics, and account for time-correlated noise in the data. We derive new secondary-eclipse depths of 0.119 $\pm$ 0.007\%, 0.123 $\pm$ 0.006\%, 0.134 $\pm$ 0.035\%, and 0.215 $\pm$ 0.008\% in the 3.6, 4.5, 5.8, and 8.0 {\micron} bandpasses, respectively. We feed these results into a Bayesian atmospheric retrieval analysis and determine that it is unnecessary to invoke a thermal inversion to explain our secondary-eclipse depths. The data are well-fitted by a temperature model that decreases monotonically between pressure levels of 1 and 0.01 bars. We conclude that there is no evidence for a thermal inversion in the atmosphere of HD~209458b.

\end{abstract}

\section{INTRODUCTION}
\label{intro}

\begin{table*}[tb]
\centering
\caption{\label{tab:ObsDates} 
Observation Information}
\begin{tabular}{ccccccccc}
    \hline
    \hline
    Label\tablenotemark{a}       
             & Wavelength   &\textit{Spitzer} Program  & Observation       & Duration  & Frame Time    & Good Frames\tablenotemark{b}  & {\em Spitzer} & Previous              \\ 
             & [{\micron}]  & (PI)                     & Start Date        & [hours]   & [seconds]     &               & Pipeline     & Publications          \\
    \hline
    Channel 1 (2005)   & 3.6          & 20532 (Charbonneau)      & November 28, 2005 & 8.1       & 0.1        & 35805          & S18.18.0     & Knutson, et al (2008) \\
    Channel 2 (2005)   & 4.5          & 20532 (Charbonneau)      & November 28, 2005 & 8.1       & 0.1        & 35800          & S18.18.0     & Knutson, et al (2008) \\
    Channel 3 (2005)   & 5.8          & 20532 (Charbonneau)      & November 28, 2005 & 8.1       & 0.1        & 35775          & S18.18.0     & Knutson, et al (2008) \\
    Channel 4 (2005)   & 8.0          & 20532 (Charbonneau)      & November 28, 2005 & 8.1       & 0.1        & 35762          & S18.18.0     & Knutson, et al (2008) \\
    Channel 4 (2007)   & 8.0          & 40280 (Knutson)          & December 24, 2007 & 8.1       & 0.4        & 87270          & S18.18.0     & -    \\
    Channel 2 (2010a)   & 4.5       & 60021 (Knutson)          & January 17, 2010  & 8.1       & 0.4        & 67960          & S18.18.0     & -    \\
    Channel 2 (2010b)   & 4.5       & 60021 (Knutson)          & January 20, 2010  & 8.2       & 0.4        & 68414          & S18.18.0     & -    \\
    Channel 1 (2011a)   & 3.6       & 60021 (Knutson)          & January 12, 2011  & 8.1       & 0.1        & 221572         & S18.18.0     & -    \\
    Channel 1 (2011b)   & 3.6       & 60021 (Knutson)          & January 15, 2011  & 8.0       & 0.1        & 220445         & S18.18.0     & -    \\
    \hline
\end{tabular}
\tablenotetext{1}{We label each dataset by its wavelength bandpass and the year the data was taken; "Channel" refers to a wavelength region or bandpass. For clarity we include the year the dataset was obtained.}
\tablenotetext{2}{There are many fewer usable frames in the 2005 datasets because \textit{Spitzer} cycled between its four IRAC detectors in order to acquire target data during a single occultation of HD~209458b.}
\end{table*}

\begin{figure}[tbh!] 
\centering
\includegraphics[width=1.0\linewidth,clip]{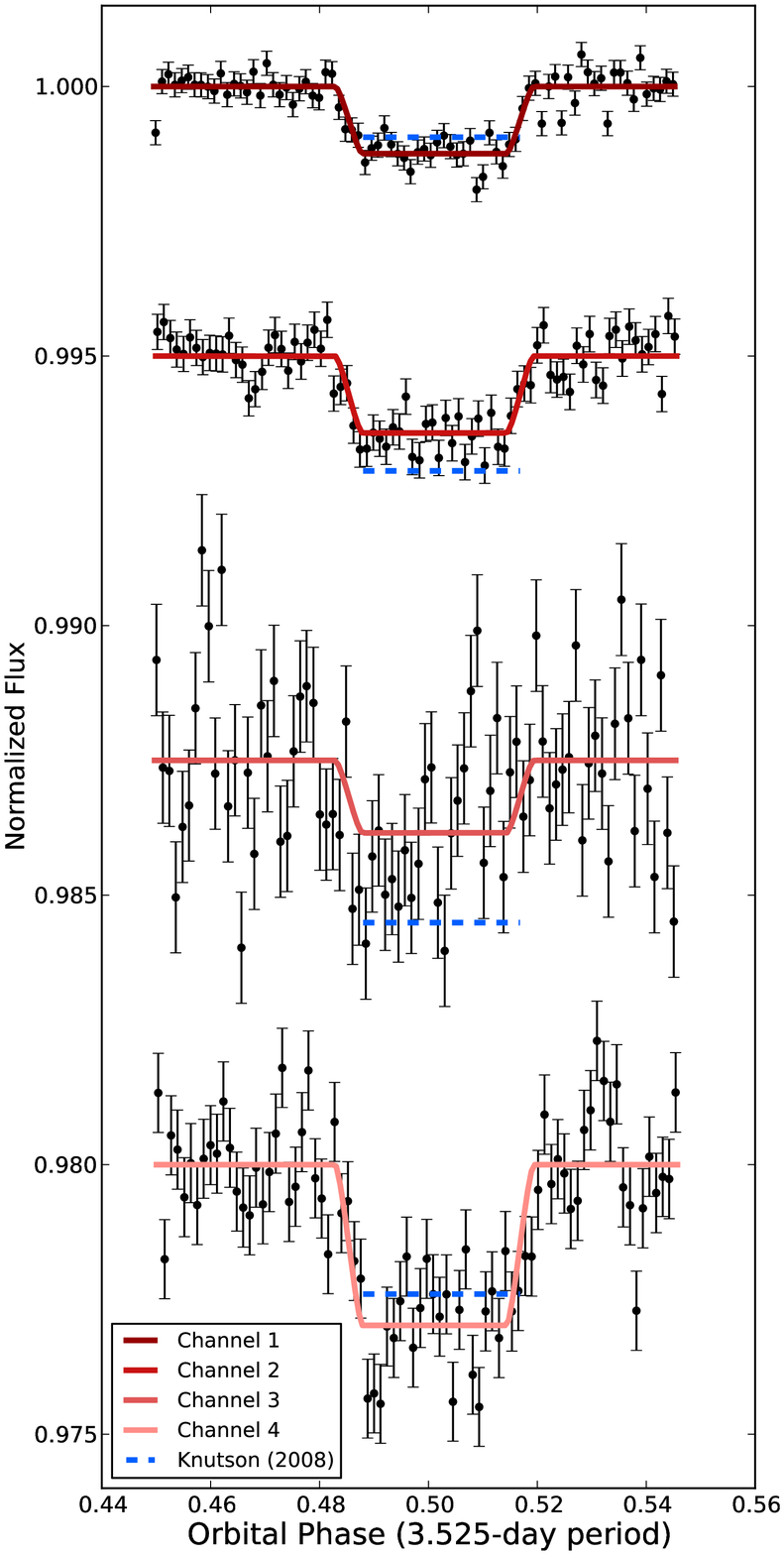}
\caption{\label{fig:FIG_2005}{
Light curves rendered from all bandpass data captured in 2005. Black points are binned data with 1$\sigma$ error bars, normalized to the system flux, and offset for ease of comparison. Colored lines in shades of red represent best fit light curves. Dashed blue lines represent eclipse depths quoted by \citet{Knutson2008}. }}
\end{figure}

As more exoplanets are discovered and studied every year, it becomes necessary to develop a comprehensive understanding of exoplanet characteristics. Characterizing the atmospheres of exoplanets is the first step to probing their interior structures and formation histories. With this in mind, an important question to ask is whether or not an exoplanet possesses a thermal inversion in its atmosphere. If an atmosphere has a large opacity at optical or ultraviolet wavelengths relative to the opacity at thermal infrared wavelengths, then the atmosphere can absorb incident stellar radiation at high altitudes without efficiently radiating this energy back to space. This would warm a region of the upper atmosphere relative to a deeper one, constituting a thermal inversion. A non-inverted atmosphere simply decreases in temperature with increasing altitude. The thermal structure of an atmosphere depends on its opacities, and therefore its composition. \\
\indent Given their size relative to their host stars, we have so far been most successful at detecting and analyzing the atmospheres of hot Jupiters. There have been notable detections of thermal inversions in several of these exoplanets, including HD~149026b \citep{Harrington2007}, HD~209458b \citep{Burrows2007, Knutson2008}, and XO-1b \citep{Machalek2008}. The diversity of hot Jupiters in which thermal inversions have been detected has led to concerted efforts to classify hot Jupiters on the basis of their atmospheric profiles, host stars, and chemical abundances \citep{Fortney2008, Knutson2010, Madhu2012}. \\
\indent All thermal inversion detections are based on data from the \textit{Spitzer Space Telescope}. Until the cryogen was depleted in 2009, \textit{Spitzer} allowed for the acquisition of photometric data in the 3.6, 4.5, 5.8, and 8.0 {\micron} bandpasses. Data from these wavelength ranges can place constraints on the atmospheric abundances of H$_2$O, CO, CO$_2$, and CH$_4$. By observing in the \textit{Spitzer} bandpasses as a planet passes behind its host star, in what is termed a secondary eclipse, it is possible to detect light directly from the planet and to determine at which wavelengths there is absorption or emission of a given chemical species. A spectral feature in absorption indicates a monotonically decreasing temperature with altitude, while a spectral feature in emission indicates a thermal inversion. \\
\indent  There has been difficulty in determining the nature of the chemical species that would account for thermal inversions. TiO and VO could exist in the gas phase at high altitudes in the atmospheres of irradiated giant planets \citep{Hubeny2003}, but there are questions as to whether these heavy absorbers could remain at high altitude in the hydrogen-dominated atmospheres of hot Jupiters \citep{Spiegel2009}. Aside from the intense vertical mixing necessary to keep TiO/VO aloft, these absorbers could be depleted by the night-side cold traps of tidally locked hot Jupiters \citep{Showman2009, Parmentier2013} or be dissociated by the intense ultraviolet radiation from the nearby host star \citep{Knutson2010}. \\
\indent One of the first secondary-eclipse observations by \textit{Spitzer} resulted in the detection of emission features in the spectrum of HD~209458b \citep{Knutson2008}. Previous models invoked a thermal inversion to fit the quoted eclipse depths \citep{Burrows2007, Madhu2009, Line2014}, but all of them rely upon four bandpass-averaged photometric points acquired in 2005, predating many systematic corrections now commonplace for \textit{Spitzer} observations. \\
\indent Recently, the thermal inversions of other exoplanets have been called into question. The detection of a thermal inversion in the atmosphere of HD~149026b rested on a single observation at 8.0 {\micron} \citep{Harrington2007}, but was refuted by \citet{Stevenson2012a} who used additional observations at more wavelengths to rule out an inversion. The proposed thermal inversion on the relatively cool hot Jupiter XO-1b may better be explained by a supersolar C/O ratio rather than an optical absorber in the atmosphere \citep{Madhu2012}. Given these findings and advancements in the field of exoplanet atmosphere characterization, we feel it is appropriate to re-examine the original data used to identify the thermal inversion in HD~209458b, as well as to analyze newer secondary-eclipse data that has since become available. We use up-to-date techniques and models to perform a complete, self-consistent analysis of the \textit{Spitzer} data, with the goal of investigating the thermal structure of HD~209458b's atmosphere. \\

\section{DATA ANALYSIS}
\label{sec:obs}

We investigate HD~209458b secondary-eclipse data acquired with \textit{Spitzer Space Telescope}'s InfraRed Array Camera \citep[IRAC,][]{IRAC}. IRAC has two kinds of detectors. The InSb detector arrays observe in the 3.6 and 4.5~{\micron} bandpasses while the Si:As detector arrays observe in the 5.8 and 8.0~{\micron} bandpasses. Until 2009, when the cryogen supply on board was depleted, this instrument was used to collect data in these four broad photometric bandpasses. The \textit{Spitzer} Warm Mission continues to observe with the InSb detector arrays in the 3.6 and 4.5~{\micron} bandpasses.\\
\indent The \textit{Spitzer} datasets that are the focus of this work each include an occultation of HD~209458b as well as out-of-eclipse baseline of the system. Data points outside of the eclipse help to constrain the instrument systematics as well as provide a comparison to the eclipse depth from which we derive constraints on the atmospheric composition. \\
\indent \textit{Spitzer}'s pipeline processes the raw detector data taken in subarray mode, removing well understood instrumental signatures, and provides basic calibrated data (BCD) files. In order to observe a 6\textsuperscript{th} K-magnitude star like HD~209458 without saturating the IRAC detectors, the observers worked in subarray mode, capturing 64 32$\times$32 pixel frames per BCD file. We then use our Photometry for Orbits, Eclipses, and Transits (POET) pipeline to further reduce the data and remove subtle instrument systematics from the light curve of the secondary eclipse \citep{Stevenson2012b, Cubillos2013}. At the start of the POET pipeline, we create a mask for bad pixels in the BCD frames and combine it with the bad pixel mask provided by \textit{Spitzer}. We then determine the 2D Gaussian center in each frame in order to perform aperture photometry. We test photometric aperture radii in 0.1 pixel increments and determine the best aperture by fitting a common model to the data at each aperture increment. We then compare the resulting standard deviation of the normalized residuals (SDNR) values. We proceed with the aperture size that produces the lowest SDNR value. \\
\indent Once we determine the best aperture size, the bulk of the work of producing a clean light curve goes into exploring and modeling the instrument systematics unique to each dataset. In all datasets we use the uniform source equations described by \citet{Mandel2002} and employ the Levenberg-Marquardt algorithm to find the best-fit results for the free parameters. We fix the parameters of secondary-eclipse duration and ingress/egress times to the calculated values of 3.08 hours and 0.42 hours, respectively. We then implement a differential-evolution Markov-Chain Monte Carlo (\mbox{DEMCMC}) with 10$^5$ steps in each of 10 chains in order to explore correlations in the parameter space and estimate uncertainties \citep{terBraak2006}.\\
\indent A key consideration is the intra-pixel effect, by which certain areas of a given pixel are more sensitive to incoming photons, and thus slight variations in stellar position translate into variations in flux that are larger than the secondary-eclipse depth we are looking for \citep{Charbonneau2005, Knutson2008}. This effect is most notable in the 3.6 and 4.5~{\micron} bandpasses, but it can also arise when working with small aperture sizes in the longer wavelength bandpasses. To model this systematic we employ a BLISS map, which uses a spline to fit the sub-pixel sensitivity at high resolution \citep{Stevenson2012a}. We fit this and other systematics simultaneously with the secondary-eclipse in order to derive our eclipse depths and uncertainties.\\
\indent We investigate all publicly available \textit{Spitzer} data to obtain our results. As outlined in Table~\ref{tab:ObsDates}, these datasets include three eclipses at 3.6 and 4.5 {\micron} (Channels 1 and 2, respectively), one eclipse at 5.8 {\micron} (Channel 3), and two eclipses at 8.0 {\micron} (Channel 4). In Section~\ref{2005data} we discuss our treatment of the original 2005 datasets (one eclipse in each channel), which yielded the result of a thermal inversion for \citet{Knutson2008}. In Section~\ref{new datasets} we discuss our treatment of more recently  acquired datasets (eclipses in Channels 1, 2, and 4), including our investigation of the presence of time-correlated noise in the data.

\subsection{Re-analysis of previously published data} 
\label{2005data}

\begin{table*}[tbh]
\centering
\caption{\label{tab:2005DataTable1} 
Comparison of derived eclipse depths from \textit{Spitzer} program 20523 (Charbonneau, PI)}
\begin{tabular}{lccccc}
    \hline
    \hline
    Label           & Wavelength  & Aperture  & Ramp                       & Intra-Pixel		  & Eclipse Depth	       \\
                    & [{\micron}] & [pixels]  & Model\tablenotemark{c}     & Sensitivity Map	& [\%]						     \\
    \hline
    Channel 1	\tablenotemark{a}      & 3.6         & 2.3       & quadratic (\ref{eqnquad})  & BLISS	          & 0.113 $\pm$ 0.010 \\  
    Channel 1 (K08)\tablenotemark{b} & 3.6         & 5.0       & quadratic                  & quadratic        & 0.094 $\pm$ 0.009 \\
    Channel 2 \tablenotemark{a}      & 4.5         & 2.1       & linear (\ref{eqnlin})      & BLISS            & 0.167 $\pm$ 0.014 \\
    Channel 2 (K08)\tablenotemark{b} & 4.5         & 5.0       & quadratic                  & quadratic        & 0.213 $\pm$ 0.015 \\
    Channel 3 \tablenotemark{a}      & 5.8         & 2.2       & quadratic (\ref{eqnquad})  & None             & 0.134 $\pm$ 0.035 \\
    Channel 3 (K08)\tablenotemark{b} & 5.8         & 3.5       & quadratic of ln            & None             & 0.301 $\pm$ 0.040 \\     
    Channel 4 \tablenotemark{a}     & 8.0         & 2.8       & quadratic (\ref{eqnquad})  & None             & 0.303 $\pm$ 0.023 \\
    Channel 4 (K08)\tablenotemark{b} & 8.0         & 3.5       & quadratic of ln            & None             & 0.240 $\pm$ 0.026 \\
    \hline
\tablenotetext{1}{Values and functions derived from this work, using the same dataset, obtained in 2005, as \citet{Knutson2008}.}
\tablenotetext{2}{Values and functions quoted from \citet{Knutson2008}.}
\tablenotetext{3}{Ramp equations used in this work are designated by number and can be referenced in section~\ref{ch1,2}.}
\end{tabular}
\end{table*}

In 2005, \textit{Spitzer} program 20523 (David Charbonneau, PI) used all four bandpasses for a duration of 8.1 hours to observe a single occultation as well as baseline of HD~209458b \citep{Knutson2008}. To keep from saturating the 3.6 {\micron} detector, the 2005 datasets were taken with 0.1 second exposure times (Table~\ref{tab:ObsDates}). Observing the same eclipse in all four IRAC bandpasses eliminates the concern of variability from one eclipse to another; however, in subarray mode, multiple detector arrays cannot acquire target data simultaneously. All four IRAC detectors collected data throughout the duration of the observation, but the telescope continuously re-pointed, cycling between the detectors such that only one acquired data of HD~209458b in eclipse at a time. Over the course of one cycle, a single detector collected 4 BCD files of target data (a $``$batch$"$) before the telescope re-pointed. While one detector was acquiring target data, the rest were collecting data from adjacent fields. Taking into account the time needed to re-point, this effectively lowers the duty cycle of each detector to less than 25\% compared to typical IRAC data. \\
\indent Due to the constant re-pointing of the telescope and the time lapses between batches of target data, systematic errors grow and information is lost, and the observational strategy employed in the 2005 HD~209458b campaign is no longer used. Since our goal is to provide a complete picture of the atmosphere of HD~209458b, we perform a thorough analysis of this dataset and compare it to subsequent datasets taken in the 3.6, 4.5, and 8.0~{\micron} bandpasses. The 2005 dataset provides our only constraints on HD~209458b's atmosphere at 5.8 {\microns}. \\
\indent In analyzing this original dataset, \citet{Knutson2008} found a systematic ramp that occurred in the first BCD file of every batch of target data taken in a given bandpass. They therefore decided to remove this first BCD file from each batch, depleting the remaining usable data in each bandpass by an additional 25\%. We apply a different approach in order to retain a maximal amount of data. By separately analyzing the first, second, third, and fourth BCD files of every batch of data in a given bandpass, we are able to investigate the systematic noise that may occur over the course of a single pointing of \textit{Spitzer}. In the 3.6, 4.5, and 8.0~{\micron} bandpasses, the first BCD files differ significantly from the others with regards to the system flux, or baseline; there is little deviation in the first BCD file of the 5.8 {\micron} bandpass. In each bandpass we perform a joint fit across the 4 BCD files of every batch, allowing only the system flux to vary from one BCD file to the next, while the rest of the model and ramp parameters are shared. With this method we are able to keep all frames available in each bandpass, while still accounting for changes in baseline flux that arise due to the constant re-pointing of \textit{Spitzer} during observation. \\
\indent The noise in the 2005 dataset at 5.8 {\micron} is comparable in size to the secondary-eclipse signal, and our ramp models have difficulty finding the ingress and egress points. We take advantage of the fact that the four datasets obtained in this observational campaign target the same eclipse. We perform a joint fit between all four bandpasses to determine a weighted average eclipse time. We then fix this eclipse time for the 5.8 {\micron} bandpass data in order to produce a light curve model. \\
\indent As a test we perform a separate analysis of the 2005 dataset with the parameters outlined by \citet{Knutson2008}. We employ similar systematic model components and aperture sizes, and we eliminate the first BCD file from every batch in a given bandpass. Even with this similar set-up of parameters, our new treatment of the data, especially with regard to our handling of systematic errors, recovers discrepant values from those reported by \citet{Knutson2008}. \\
\indent The difference between our best results for the 2005 dataset and those reported by \citet{Knutson2008} is over 1$\sigma$ at 3.6 {\micron}, over 2$\sigma$ at 4.5 {\micron}, over 3$\sigma$ at 5.8 {\micron}, and almost 2$\sigma$ at 8.0 {\micron}. Figure~\ref{fig:FIG_2005} and Table~\ref{tab:2005DataTable1} illustrate these discrepancies in eclipse depth. These differences may be due in part to the fact that we received our data from \textit{Spitzer} pipeline version S18.18.0, while \citet{Knutson2008} received it from version S13.0. Moreover, our data reduction techniques employ current methodology, especially with regards to mapping intra-pixel sensitivities in the 3.6 and 4.5 {\micron} bandpasses. \\

\subsubsection{3.6 and 4.5 $\mu$m bandpasses}
\label{ch1,2}

\begin{figure}[tb]
\centering
\includegraphics[width=1.0\linewidth,clip]{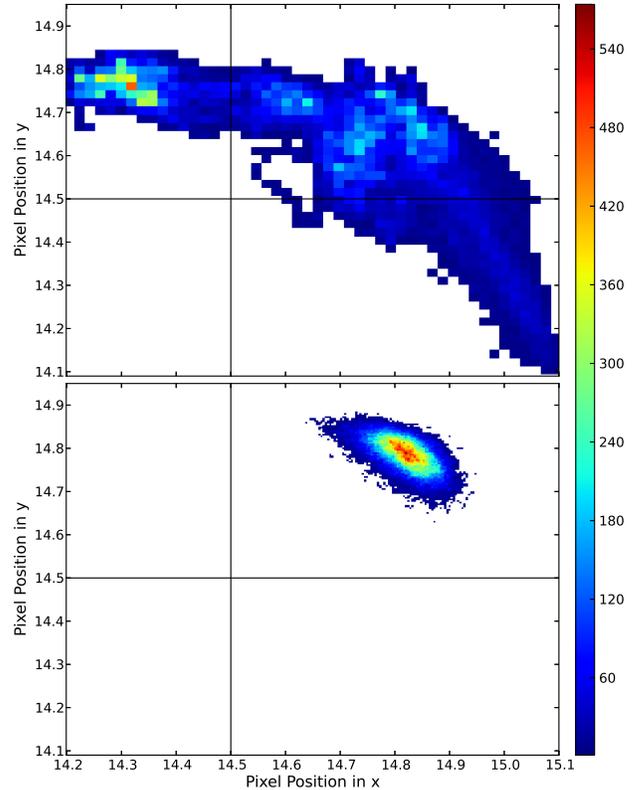}
\caption{\label{BLISShistogram}{
2D BLISS map histograms for Channel 1 (2005) and Channel 1 (2011a), top and bottom, respectively. Axes represent spatial locations on a detector pixel, while the apparent pixelation in the figures represents bin size. The color bar to the right of the histograms indicates the number of frames per bin. The black lines are the boundaries of a detector pixel. We compare the quality of BLISS mapping between the Channel 1 (2005) and Channel 1 (2011a) datasets. The BLISS map of the Channel 1 (2005) is smeared along the detector with bin size of 0.019 pixels in length and width. The Channel 1 (2011a) dataset has more than 6 times the amount of data (Table~\ref{tab:ObsDates}), all of which are concentrated within a fraction of a detector pixel. The resulting BLISS map is much more comprehensive, with comparatively small bin sizes of 0.004 pixels in length and width.}}
\end{figure}

\indent The 3.6 and 4.5~{\micron} bandpasses exhibit significant intra-pixel variability. In both bandpasses we have few good frames relative to subsequent datasets (Table~\ref{tab:ObsDates}), meaning that our measured eclipse depths for these datasets have relatively high uncertainties. The BLISS map employed in this analysis relies on having enough photons received by a given sub-pixel region to map its sensitivity. Looking at a 2D histogram of the data in Figure \ref{BLISShistogram}, it is clear that there are not enough photons clustered on the same pixel to do this effectively, and so the map becomes highly flexible with large uncertainty. We attempted to use a quadratic function in both the x- and y-positions to model intra-pixel sensitivity, but while this method can fit the overall ramp, in many cases it was unable to detect the occultation. \\
\indent At 3.6~{\micron} we find that the best aperture has a radius of 2.3 pixels with a time-dependent quadratic function of
\begin{equation} 
\label{eqnquad} 
R( t) = 1 + r\sb{2}(t - 0.5) + r\sb{3}(t - 0.5)^2
\end{equation}
as a best fit for the ramp in the data, where $t$ is time in units of phase, and $r_{2}$ and $r_{3}$ are free parameters. With these parameters we determine an eclipse depth of 0.113 $\pm$ 0.010\% (Figure~\ref{fig:FIG_2005}, Table~\ref{tab:2005DataTable1}). At 4.5~{\micron} we use an aperture radius of 2.1 pixels and employ a linear function
\begin{equation}
\label{eqnlin}
R( t) = 1 + r\sb{2}(t - 0.5)
\end{equation}
as a best fit for the ramp. Here we determine an eclipse depth of 0.167 $\pm$ 0.014\% (Figure~\ref{fig:FIG_2005}, Table~\ref{tab:2005DataTable1}).\\
\indent Given the difficulties in developing an effective pixel map for these bandpasses we turn to other datasets taken in 2010 and 2011 to further explore the parameter space and to obtain more accurate eclipse depths and uncertainties (Section~\ref{new datasets}).\\

\begin{table*}[tb]
\centering
\caption{\label{tab:SubDataTable} 
Derived eclipse depths from \textit{Spitzer} programs 40280, 60021 (Knutson, PI)}
\begin{tabular}{ccccccc}
    \hline
    \hline
    Label     & Wavelength	& Aperture Size  & Ramp      & Intra-Pixel		    & $\gamma$  & Eclipse Depth	     \\
              & [{\micron}]	& [pixels]       & Model     & Sensitivity Map	  &           & [\%]						    \\
    \hline
    Channel 1 (2011a) 		& 3.6		      & 2.3		         & quadratic & BLISS	      & 0.69      & 0.119 $\pm$ 0.007	  \\
    Channel 1 (2011b) 	  & 3.6	    	  & 2.6		         & quadratic & BLISS        & 1.70      & 0.105 $\pm$ 0.011   \\
    Channel 2 (2010a)    & 4.5         & 2.8            & linear    & BLISS        & 1.83      & 0.132 $\pm$ 0.005   \\
    Channel 2 (2010b)    & 4.5         & 3.0            & quadratic & BLISS        & 0.95      & 0.123 $\pm$ 0.006   \\
    Channel 4 (2007)       & 8.0         & 3.6            & quadratic & BLISS        & -         & 0.215 $\pm$ 0.008   \\
    \hline
\end{tabular}
\end{table*}

\subsubsection{5.8 and 8.0 $\mu$m bandpasses}
\label{ch3,4}

\indent The 5.8 and 8.0~{\micron} bandpasses do not exhibit intra-pixel sensitivity. We attempted to use the BLISS map in both bandpasses since the small aperture sizes we use can lead to greater pixelation effects. In neither bandpass were we able to find a sub-pixel bin size at which the BLISS map outperformed a nearest-neighbor interpolation \citep{Stevenson2012a}. We ultimately achieved consistent results without using the BLISS map and we do not include it in our final analysis of these bandpasses. \\
\indent At 5.8 and 8.0 ~{\micron} we use aperture sizes of 2.2 and 2.8 pixels, respectively, and a quadratic ramp to model the data. At 5.8 ~{\micron} we achieve an eclipse depth of 0.134 $\pm$ 0.035\%, and at 8.0 ~{\micron} we achieve an eclipse depth of 0.303 $\pm$ 0.023\% (Table~\ref{tab:2005DataTable1}, Figure~\ref{fig:FIG_2005}). While we have no further data at 5.8~{\micron} against which to compare our results, we do look at a subsequent dataset in the 8.0~{\micron} bandpass, taken in 2007 (Section~\ref{new datasets}). \\

\begin{figure}[tb]
\centering
\includegraphics[width=1.0\linewidth,clip]{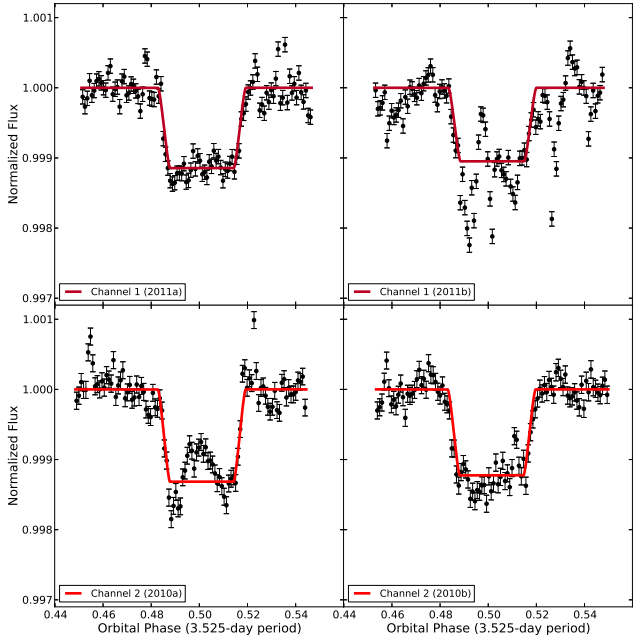}
\caption{\label{fig:Ch1,Ch2_lightcurves}{
Light curves rendered from the Channel 1 (2011a), Channel 1 (2011b), Channel 2 (2010a), and Channel 2 (2010b) datasets. The secondary eclipses in each bandpass are consecutive and were obtained by the same \textit{Spitzer} program. The Channel 1 (2011b) and Channel 2 (2010a) light curves exhibit abundant time-correlated red noise and we do not include the eclipse depths measured from these light curves in our atmospheric retrieval.}}
\end{figure}

\subsection{Analysis of newer datasets} 
\label{new datasets}

\begin{figure}[tb]
\centering
\includegraphics[width=1.0\linewidth,clip]{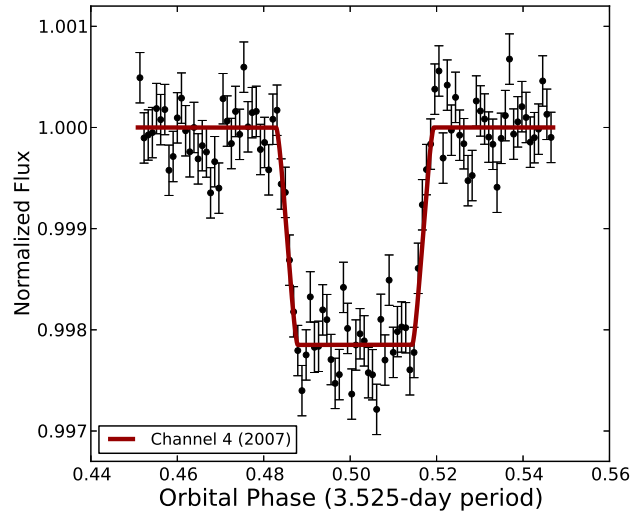}
\caption{\label{fig:Ch4_2007}{
Light curve rendered from the Channel 4 (2007) dataset. Black points are binned data with 1$\sigma$ error bars. Red line indicates best fit light curve model.}}
\end{figure}

In 2007, \textit{Spitzer} program 40280 (Heather Knutson, PI) took a half-orbit phase curve of HD~209458b in the 8.0 ~{\micron} bandpass (Table~\ref{tab:ObsDates}). In 2010 and 2011, \textit{Spitzer} acquired data of HD~209458b as part of a campaign to capture a complete phase curve, this time in the 4.5 and 3.6 ~{\micron} bandpasses, respectively (\textit{Spitzer} program 60021, Heather Knutson, PI). From these observations we extract one occultation of HD~209458b in the 8.0~{\micron} bandpass, and two consecutive occultations of HD~209458b in each of the 3.6 and 4.5~{\micron} bandpasses; however, only one eclipse in each of the lower two bandpasses is $``$clean,$"$ while the other exhibits abundant noise 
(Figure \ref{fig:Ch1,Ch2_lightcurves}). It is unclear whether the abundant noise in these eclipses is a result of short-timescale stellar activity or unknown instrument systematics. Given that HD~209458 is a relatively inactive star \citep{Knutson2010}, we determine that these particular datasets are plagued by time-correlated noise that does not manifest predictably. \\
\indent We clip the phase curve data around the secondary-eclipses to prevent the periodic flux variations from affecting the measured secondary-eclipse depths. Clipping the phase curve to just the secondary eclipses plus a few hours of baseline means that we can use simple linear or quadratic equations to fit the temporal systematics in the data. \\
\indent In performing aperture photometry we again choose aperture sizes that produce the lowest SDNR values. The best apertures tend to be small ($\leq$ 3.6 pixel radii), but smaller apertures are more susceptible to flux variations from imprecise stellar centering. We therefore use Time-series Image Denoising (TIDe) to remove high-frequency jitter in the stellar centering while performing photometry on the unfiltered images \citep{Stevenson2012b}. The temporal continuity of the data and short exposure times for each frame allow us to employ this method. The 3.6~{\micron} bandpass data were captured with 0.1 second exposures, and we see a decrease in SDNR when TIDe is used. The 4.5 and 8.0~{\micron} bandpass data were captured with 0.4 second exposures, and thus we see only a slight decrease in SDNR and almost no change in eclipse depth when we apply TIDe. \\
\indent Due to the continuity and equal spacing of the data, we are able to better estimate our uncertainties by taking into account the contribution of unknown time correlations in the data, or red noise \citep{Carter2009b}. Investigating the red noise does not change our eclipse depths, but it does ensure that our uncertainties are large enough to account for these time correlations. \\
\indent The red noise follows a power spectral density expressed by the equation $1/f^{\gamma}$ \citep{Carter2009b}. We explore all available wavelets in the Python PyWavelets package and choose the best one, based on SDNR and Bayesian information criterion (BIC) values, to model the time-correlated noise in the data. We determine the coefficients for white noise, red noise, and $\gamma$ by incorporating these parameters into the \mbox{DEMCMC} used to fit the eclipse parameters, temporal systematics, and BLISS map. A result of $\gamma = 1$ implies that the noise in the data is made up of equal parts uncorrelated white noise ($\gamma = 0$) and time-correlated red noise ($\gamma = 2$). Assuming a $\gamma$ of 1, as is done in the example provided by \citet{Carter2009b}, can result in an under- or overestimation of the uncertainties, depending on the amount of correlated noise. Freeing $\gamma$ constrains the relative amounts of white and red noise present in the data and allows us to more accurately predict uncertainties for our eclipse depths. \\
\begin{table}[tb]
\centering
\caption{\label{tab:SummaryBestData}
Best eclipse depths, used in retrieval (Fig. \ref{fig:spectrum_tpprofile})}
\begin{tabular}{cccc}
    \hline
    \hline
    Label    & Wavelencth    & Eclipse Depth             & Spitzer Program       \\
             & [{\micron}]   & [\%]                      & (PI)                  \\
    \hline
    Channel 1 (2011a)   & 3.6           & 0.119 $\pm$ 0.007	    & 60021 (Knutson)       \\
    Channel 2 (2010b)   & 4.5           & 0.123 $\pm$ 0.006      & 60021 (Knutson)       \\
    Channel 3 (2005)      & 5.8           & 0.134 $\pm$ 0.035      & 20523 (Charbonneau)   \\
    Channel 4 (2007)      & 8.0           & 0.215 $\pm$ 0.008      & 40280 (Knutson)       \\
    \hline
\end{tabular}
\end{table}
\indent As an experiment, we fix $\gamma$ to several values between 0.0 and 2.0 for the Channel 1 (2011a) dataset. We find an increase in uncertainty with increasing $\gamma$. Had we fixed $\gamma$ to 1 we would have overestimated our eclipse depth uncertainty for this dataset by 30\%. From looking at Table~\ref{tab:SubDataTable} it is clear that the $``$noisy$"$ eclipses have $\gamma$ values greater than 1, implying that there is more red noise than white noise in these datasets. In the case of the Channel 4 (2007) dataset, the $\gamma$ parameter is driven so low that we conclude that there is no discernible time-correlated noise in the data. We plot the normalized RMS residuals versus bin size and verify that they follow the predicted standard error for Gaussian noise \citep{Pont2006}. \\
\indent Once we perform photometry, we fit the eclipse parameters, temporal systematics, BLISS map, white noise, red noise, and gamma value simultaneously to determine our final results and uncertainties. At 3.6~{\micron} we use an aperture of 2.3 pixels and a quadratic ramp to model the data, and we achieve an eclipse depth of 0.119 $\pm$ 0.007\%. At 4.5~{\micron} we use an aperture of 3.0 pixels and a linear ramp to achieve an eclipse depth of 0.123 $\pm$ 0.006\%. We are able to achieve a good fit and comparable eclipse depths to within 1$\sigma$ in the 8.0~{\micron} channel, with or without a BLISS map (Figure \ref{fig:Ch4_2007}). We choose to include the BLISS map, and for our final analysis we use an aperture of 3.6 pixels and a quadratic ramp to achieve an eclipse depth of 0.215 $\pm$\ 0.008\% (Table~\ref{tab:SubDataTable}). \\
\indent As a final check for our measured eclipse-depth uncertainties, we inject 30 fake transit signals into out-of-eclipse baseline regions of the Channel 1 (2011) phase curve, prior to applying any of our models or the BLISS map. Upon successfully retrieving all of the light curves, we find the mean and distribution of eclipse depths to be consistent with our reported best-fit depth and uncertainty. \\

\section{ATMOSPHERIC RESULTS} 
\label{sec:results}

\begin{figure*}[tb]
\centering
\includegraphics[width=1.0\linewidth,clip]{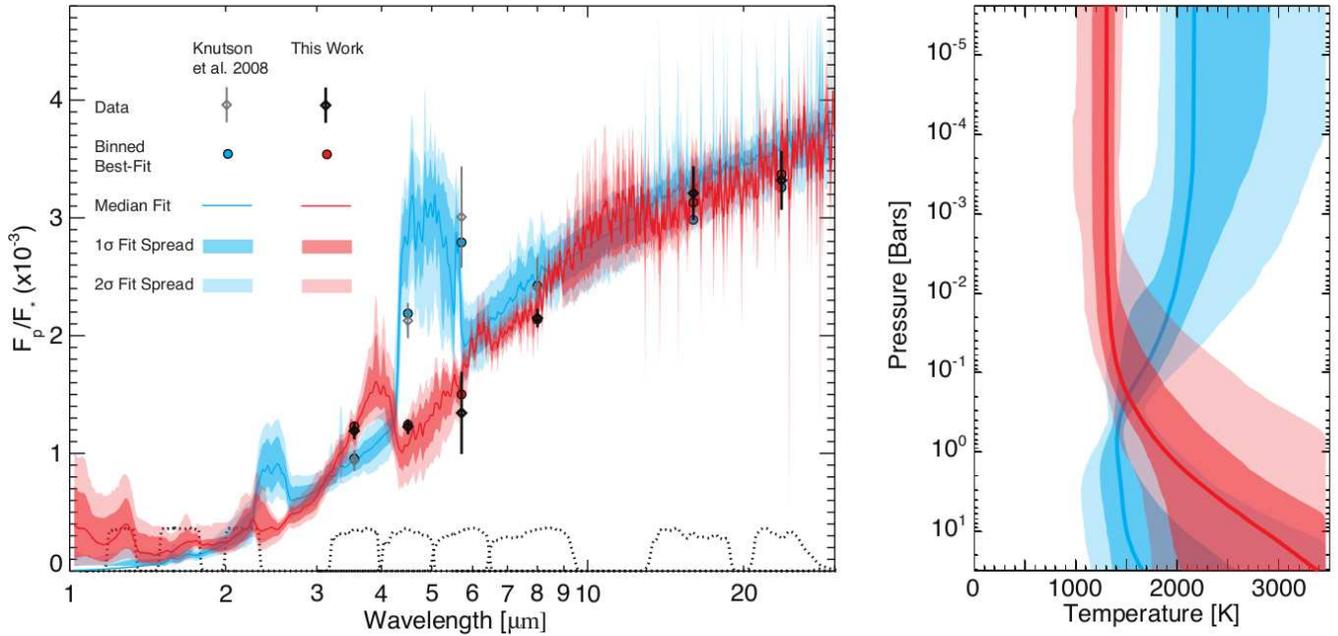}
\caption{\label{fig:spectrum_tpprofile}{
Summary of the atmospheric retrieval results for our data and for the Knutson et al. (2008) data. The left panel compares model atmospheres based on retrievals of our best eclipse depths (red) and of eclipse depths presented by \citet[blue]{Knutson2008}. Shading indicates 1$\sigma$ and 2$\sigma$ confidence intervals. Bandpass averages are given in red and blue circles at all four IRAC bandpasses (3.6, 4.5, 5.8, and 8.0 {\micron}), as well as at 16~{\micron} and 24~{\micron}. We take the points representing the latter two bandpasses from the literature \citep{Swain2008, Crossfield2012a}. The four black diamonds with error bars represent our best eclipse depths in each bandpass (Table~\ref{tab:SummaryBestData}), while the grey diamonds with error bars represent the quoted eclipse depths from \citet{Knutson2008}. Dashed lines at the bottom of the figure indicate bandpass coverage for the J, H, and K bands, as well as the \textit{Spitzer} IRAC, IRS, and MIPS instruments. The bandpass averages fall within error of our results. We are able to achieve smaller error bars in the 3.6, 4.5, and 8.0 {\micron} bandpasses because we use observations that contain many more data points and data reduction techniques that more accurately account for stellar centroid variations, intra-pixel sensitivity, and time-correlated noise. The panel on the right shows the temperature-pressure profiles of HD~209458b based on a retrieval of our best eclipse depths (red) and the eclipse depths quoted by \citet[blue]{Knutson2008}. Using the quoted eclipse depths we are able to reproduce the thermal inversion discussed by \citet{Knutson2008}, but we do not find this inversion when we perform a retrieval of our measured eclipse depths. We are sensitive to atmospheric pressures of 1 to 0.01 bars, and at these pressures we are able to tightly constrain the thermal profile such that there is little overlap between our confidence intervals (shaded regions) and those of the reproduced literature thermal profile. The sections of the temperature-pressure profiles that fall outside of this pressure region are extrapolated from the temperature profile parameterization.}}
\end{figure*}

In order to characterize the atmosphere of HD~209458b, we perform a Bayesian retrieval analysis of photometric data resulting from our best light curves (Table~\ref{tab:SummaryBestData}). We choose to perform our retrieval on only these results because they are least affected by correlated noise or, in the case of the 5.8~{\micron} channel, because it is the only available light curve. We do not include our other light curves from the 2005 dataset because they were captured in a sub-optimal observation mode and have large uncertainties. \\
\indent For our retrieval analysis we use the CHIMERA suite \citep{Line2013a, Line2013b, Line2014}. Briefly, CHIMERA uses three retrieval approaches to determine the allowed range of temperature profiles and abundances that are consistent with the data. We summarize the results from the \mbox{DEMCMC} approach, which is the most comprehensive of the three CHIMERA algorithms. We use a parameterized temperature profile based off of an analytic gray radiative equilibrium solution \citep[e.g.,][]{Guillot2010, Robinson&Catling2012} and four molecular absorbers, H$_2$O, CH$_4$, CO, and CO$_2$. Further details on the opacity databases and atmospheric parameterization can be found in \citet{Line2013a}. \\
\indent We compare our results to those derived by \citet{Line2014}, who performed a retrieval analysis of the eclipse depths reported by \citet{Knutson2008}. Figure \ref{fig:spectrum_tpprofile} displays the median and the 1$\sigma$ and 2$\sigma$ spreads in the spectral fits to our best secondary-eclipse depth measurements (red) as compared to those from \citet[blue]{Line2014}. Using the \citet{Knutson2008} eclipse depths, \citet{Madhu2009} and \citet{Line2014} were able to confirm the presence of a thermal inversion; however, when applying the retrieval analysis to our eclipse depths, we find no evidence for a thermal inversion at the pressure regions probed by our observations. We stress, though, that we have little sensitivity at altitudes above the \sim10 mbar level, so it is possible that a weak inversion may persist at these high altitudes with minimal impact on our data. \\
\indent We find that volumetric mixing ratio for water is well-bounded from $\sim7\times10^{-6}$ - $5\times10^{-4}$. This is in stark contrast to the water abundance found with the inversion by \citet{Line2014} where they were only able to achieve an upper limit of ~1$\times10^{-6}$ for water. The inversion solution derived from the \citet{Knutson2008} eclipse depths primarily results from the high degree of flux in the 4.5 and 5.8~{\micron} bandpasses. These two photometric bandpasses overlap with both CO and CO$_2$ absorption. The abundance of CO, which absorbs more strongly at 5.8~{\micron} than does CO$_2$, must be driven high in order to match these points. The discrepancy between the two solutions is due to the need to suppress the water abundance in the inversion model in order to prevent strong emission in the longer wavelength bandpasses. \\
\indent We also find an upper limit on the volumetric mixing ratio of methane of $\sim1\times10^{-7}$ from this analysis, consistent with \citet{Line2014}. Our results suggest a much lower abundance of CO than reported by \citet{Line2014}, again because in the inversion scenario the CO abundance must be driven to an unphysically high abundance in order to produce strong emission at 5.8~{\micron}. All of our volumetric mixing ratios are consistent, to within 1$\sigma$, with a solar-composition thermochemical equilibrium atmosphere.\\
\indent \citet{Zellem2014} analyzed the full phase curve of HD~209458b taken in the 4.5{\micron} bandpass in 2010 (\textit{Spitzer} program 60021, Heather Knutson, PI). Although discrepant to our 4.5{\micron} final result by 1.7$\sigma$, their secondary-eclipse depth measurement, reported from a full phase curve analysis, is consistent with our interpretation of a non-inverted atmosphere.

\section{CONCLUSIONS}
\label{sec:concl}

This analysis of historical \textit{Spitzer} data serves the dual purpose of exploring the systematics in the data using up-to-date techniques, as well as providing an interpretation of HD~209458b's atmosphere that is consistent with current methods. HD~209458b is known to be the prototypical exoplanet for atmospheric thermal inversions; however, the results of our analysis do not corroborate this claim. Our best fit atmospheric models do not require any species in emission at the pressures probed by the observations in order to explain our measured eclipse depths. This suggests a temperature profile that decreases with increasing altitude. \\
\indent At the time of writing there are several new observational campaigns of HD~209458b with warm \textit{Spitzer} (program 90186, PI Kamen Todorov; program 10103, PI Nikole Lewis). We look forward to comparing the results of these new observations with the ones we have derived from the historical data. Ultimately we look to future spectroscopic observations, from Wide Field Camera 3 on the \textit{Hubble Space Telescope} or from the heavily anticipated \textit{James Webb Space Telescope}, to provide data that can better constrain our models, and thereby definitively characterize the thermal profile and atmospheric composition of HD~209458b. \\

\acknowledgments
We thank Dorian Abbot for his thoughtful comments. We also thank contributors to SciPy, Matplotlib, the Python Programming Language, and the free and open-source community. This research has made use of the NASA/IPAC Infrared Science Archive, which is operated by the Jet Propulsion Laboratory, California Institute of Technology, under contract with the National Aeronautics and Space Administration (NASA). Funding for this work has been provided by NASA grant NNX13AJ16G. J.L.B. acknowledges support from the Alfred P. Sloan Foundation. K.B.S. performed work in part during a Sagan Fellowship supported by NASA and administered by the NASA Exoplanet Science Institute (NExScI).

\bibliography{ms}

\end{document}